\ifx\mnmacrosloaded\undefined \input mn\fi
 
\def\etal{{\it et al\/}}
\Referee
%\Autonumber
\begintopmatter

\title{An adaptive multigrid solver for high-resolution
	cosmological simulations}
\author{I.~Suisalu$^{1,2}$ and E.~Saar$^2$}
\affiliation{$^1$ Royal Observatory, Edinburgh, Blackford Hill, Edinburgh EH9
3HJ,
	Scotland}
\affiliation{$^2$ Tartu Astrophysical Observatory, T\~oravere,
	EE-2444, Estonia}

\shortauthor{I.~Suisalu and E.~Saar}
\shorttitle{Adaptive multigrid solver}

\abstract{
We have developed an
adaptive multigrid code for solving the Poisson
equation in gravitational simulations.
Finer rectangular subgrids are adaptively
created in locations where the density exceeds a local
level-dependent threshold. We describe the code,
test it in cosmological simulations, and apply it
to the study of the birth and evolution of a typical
pancake singularity. The initial conditions for the
pancake are generated on the basis of the theory of
Lagrangian singularities; we follow its evolution
for a few collapse times, finding a rich substructure
in the final object.
We achieve a spatial resolution of 1/1024 of the
size of the overall computational cube
in the central parts of the pancake, with
computing time comparable to that
of the FFT-solvers.}


\maketitle

\section{Introduction}

It is common knowledge that the dynamical and spatial
resolution of standard cosmological simulations of
smooth (dark) matter evolution is rather low.
The basic particle-mesh (PM) code smooths the density
on a grid and finds the potential using FFT-type
techniques.
As shown by Bouchet, Adam \& Pellat (1985), grid
effects damp the growth of structure for all scales smaller
than at least 6 cell sizes.  This restricts
considerably the range of
scales where we can believe our models.

In order to improve the situation several
different methods have been proposed.
The best known of them is the
PPPM (particle-particle particle-mesh) code, developed
by Eastwood \& Hockney (1974) and described in detail in
their book (Hockney \& Eastwood 1981). It finds the
large-scale forces from the smoothed density, as does the basic
PM-code,
but uses the nearby mass points to calculate the detailed
short-range force. This code was adapted
for cosmological simulations by
Efstathiou, Davis, Frenk \& White (1985), and is presently
the standard code for cosmology. Couchman (1991) has
modified it for situations where the density range becomes
large and most of the time in the PPPM-code would be
spent on calculating pair interactions, introducing
adaptive mesh refinement in order to reduce the amount
of calculations of pairwise forces.

The ultimate high-resolution code is the tree code,
proposed by Barnes \& Hut (1986) and first used for
cosmological simulations by Bouchet \& Hernquist (1988).
This code treates all interactions between gravitating
masses basically on a pairwise basis and is thus free from
any grid effects.

Both the PPPM code and the tree code are based on the paradigm
of individual clouds of matter. Although a softened force
is usually used, it is not clear how well the massive clouds
represent the essentially continuous distribution of dark matter.
Thus it is necessary to develop a high-resolution code for
this class of problems.

The first of such codes, a multiple mesh particle
code was proposed by Chan, Chau, Jessop \& Jorgensen (1986), who
use local higher resolution grid patches in high density regions.
Another mesh refinement method has been proposed by
Villumsen (1989). These methods differ mainly by their approach to
solving field equations on local grids.
Chan \etal (1986) obtain the
local grid  potential as a solution to a boundary value problem for
the Poisson equation on the
local grid, where the boundary values are defined
from the solution on a global coarser grid. They find
the solution
on local grids by an efficient iterative scheme.
Villumsen (1989) considers
the local potential as
a sum of a solution for the distribution of matter on the subgrid,
using isolated boundary conditions, and the external global solution for
the full matter distribution (without the mass at refined regions).
He uses FFT to find the potentials, and special tricks to
speed up the solution for the local isolated local patches.

As we see, in the latter approach
there is no backreaction
from submesh particles to the global grid, the coarser periodical
part of the field is computed without subgrid particles.
In Chan \etal~method this backreaction is included.

An important issue intrinsic to all multiple resolution
mesh schemes is how well they treat particles that
enter subgrids. In principle, when switching to a higher resolution
force anisotropies and radial errors should be less than these on
a coarser grid, and additional errors will be generated mainly by solution
errors near grid interfaces. In case of a boundary value problem,
(Chan \etal~approach), the error depends on interpolation errors
on the boundary and on the density estimate at the grid point
next to the boundary. For the Villumsen code the errors are
generated by an external lower resolution solution from a coarser grid
and by the density estimation both at the grid boundary and at neighbouring
grid points.

In the Couchman's $AP^{3}M$ code that uses
force splitting to long-range and short-range parts
(intrinsic to $P^{3}M$), the
mesh part of the force is approximated by a smooth reference
interparticle force $R(r,a_{m})$, where $a_{m}$ is the softening scale
appropriate for the current mesh resolution.
The potential is calculated using a
modified Green`s function which minimizes
the differences between computed and reference forces.
The total mesh part of the force acting on a particle is
found by summing the force $R(r,a_{m})$ from the base (periodical)
mesh and the forces due to particles
on a refinement, calculated with a different softening scale $a_{s}$,
and the system is locally considered as isolated.

The main difference between the Chan \etal, Villumsen's and Couchman's codes
is that in the former two the aim of the mesh refinement is to
enhance the resolution of the force field through higher resolution
potential field whilst in the last code the refined grid is introduced
mainly to reduce the computational overhead in short-range force
calculation in case of heavy clustering of particles. A smaller
gridsize allows to use a smaller neighbouring radius $r_{c}$, reducing the
number of particles which should be used for the short range force
correction. Another advantage of $AP^{3}M$ is the adaptive dynamic
creation of submeshes.

This paper presents another adaptive algorithm for
a continous density distribution, based on the well-known
multigrid method (Brandt 1977). We shall give a short overview of the
method below. The main positive features of the
multigrid method are that for the case of the Poisson equation
its computational complexity scales as $O(N)$, it allows one
to control the errors inherent in the
solution of the Poisson equation, and it
lends itself naturally to adaptive refinement.
It is also very flexible in applying different boundary conditions and
allowing special treatment if necessary.

Compared with the methods described above, our implementation of the
adaptive multigrid solver
for Poisson equation is physically closest to that of
Chan \etal, as it
finds the local grid potential as the solution for the
boundary value problem defined by a coarser grid solution.
The main difference is that in the adaptive multigrid method the
creation of subgrids is a natural part of the solution process.
The local refinements are introduced during multigrid
iteration in locations where predefined error estimates demand it.
There is a tight interference between the coarse grid
solution and a finer grid solution as the coarse grid is used for
correction of the solution on the finer grid and vice-versa.
At the internal boundaries (subgrid interfaces) the values of
the potential change during the search for an overall solution,
due to changes in the
finer grid solution. This is
natural, because they should agree despite of different scales of the
discretisation and interpolation errors. The point is that
the local grid problem should be solved together with the global problem.
If we treat it separately,
we contradict with global boundary conditions.
Natural boundary
conditions for the local
problem are the Dirichlet' conditions,
as this guarantees
that the potential will be continuous at subgrid boundaries.

Comparison of our approach with the Couchman code is not so straightforward
due to different
approaches in force calculation.
There the global (periodic) part of the potential is calculated coarsely and
the
local (nonperiodic) part is found using isolated boundary conditions,
so there is no backreaction to large scales. However,
as the force resolution is enhanced
further using local direct summation,
the backreaction effects could be less inportant.

Our code has already been applied to the modelling
of gravitational microlensing (Suisalu 1991).
We describe here its application to cosmological
situations, and follow as an example the birth and
the subsequent evolution of a typical pancake singularity.

\section{Multigrid description}

Multigrid methods were introduced for solving
boundary value problems by Brandt (1977).
Quite comprehensive reviews of multigrid methods
can be found in Press \etal.\ (1992)
and Brandt (1984).

As the name implies, in this method the problem is solved
iteratively, using several sets of grids with different fineness.
In the standard case the mesh size ratio of
the grids at neighbouring levels $H/h=2$ and the points of
the coarser grid coincide with every second point of the
finer grid. The solutions on the coarser grids are used
to estimate the smooth components of the final solution and the
increasingly finer grids are used to determine details of
the solution. We refer readers to the recent introduction
in Press et~al.\ (1992, sec.~19.6)
and will not give the details of the multigrid method here,
noting here only
that we use the Full Multigrid algorithm (FMG), the Gauss-Seidel
red-black
relaxation to smooth the high-frequency error
components, and the
Full Approximation Storage (FAS) algorithm in order to
be able to adaptively introduce finer grid levels.

In standard multigrid problems the source term (r.h.s. of
the equations) is usually well determined. In gravitational
simulations, however, the density distribution is sampled
by discrete mass points, and the problem of determining
the matter density on different grid levels is rather
complicated. We shall describe our approach below.

We start with a fixed number of levels of uniform grids $G_h$ which
cover all the computing domain. On the finest level we
compute the density $D_h$ from particles using the
cloud-in-cell (CIC) algorithm in the chosen computational volume
and make a linked list of particles
inside the volume. At the end of the density computation we
find also the boundary values of the potential field $f_h$ using
this density. In this way we have defined the r.h.s. term
of our differential equation for a grid with mesh size $h$:
$$
	{\cal L}_h F_h = D_h,
$$
subject to the boundary condition
$$
	F_h = f_h,
$$
on the boundary of $G_h$, where ${\cal L}_h$ is in our case
the usual 7-point finite difference operator for the Laplace
equation.

Having solved this equation approximately on the level $G_h$,
using coarser levels to accelerate the solution process, we
proceed by adaptively introducing finer grid levels. We have used
the value of the local density as the criterion for deciding
if there was a need for finer grids, but there are other
possibilities. One of the most popular criteria used is
the local truncation
error ( the difference between the discrete ${\cal L}_h F_h$
and the analytical ${\cal L} F$) which is a natural byproduct
of any multigrid solution.
To apply the local density criteria we mark during the density
computation the gridpoints where the density is higher than a
chosen threshold. By analyzing this flag field we define the
flag clusters which could be bounded by rectangular grid
boxes. As we use at present only non-overlapping
grids some postprocessing is necessary.
\beginfigure{1}
	\vskip 6.5cm
	\caption{{\bf Figure 1.}
The forces obtained from the multigrid simulation
compared to the exact $r^{-2}$ force, versus distance
(in grid cell units). The force scale is given in decimal
logarithm, but the distance is in base two logarithm.
The exact force (labeled TH) is shown by a
solid line, the basic mean multigrid force (MG) by a dashed
line, and the mean force obtained from a two level finer adaptive
solution (AMG) by a dotted line. The two latter forces were
found by choosing randomly a position of a massive point
and calculating then forces at a number of positions around
it.
The lines labeled RF and TF show the rms relative radial and
tangential force fluctuations with respect to the mean radial
force (AMG) obtained with adpative grids.}
\endfigure

After allocating new grids, we compute the
density on these grids using the particle lists from the coarser level that
contains the current grid. We also create a new point list for this particular
grid, consisting of those points only that are inside the grid, and we
subtract this list from the pointlist of the coarser grid
in order to keep points divided between grids.
This, firstly, eliminates unnecessary scans over the full particle array,
and, secondly,
these lists are needed later on if we start to move particles around
on our collection of grids.

Besides the density we interpolate the boundary values for the potential
from those on the coarse grid in order
to specify the local Dirichlet problem. We will repeat
the multigrid solution process on new grids using coarser
grids and generating new finer grids until the desired resolution is
achieved. For every iteration we update the density on coarser
levels by the fine level density in order to keep our differential equation
consistent on different grid levels. It is possible to use the
CIC scheme for density computation, but as the cells at different
grid levels cover different volumes, the density estimates
differ for the same point at different
levels. This introduces additional noise that makes the convergence
more difficult. Thus one has either to introduce  spatial
averaging for coarser grids or to
invent a density measure which gives the same value
at the same point in space independent of the mesh size
of a grid. In the present code we have used
the full weighting scheme that finds
densities on coarser grids by averaging it
over neighbouring points on the finer
grid.

The multigrid solution process stops after making a few additional
iterations on the finest level and checking that the changes in the solution
are less than the truncation error norm on that
level times a small factor (usually 0.01).  This assures that the
discrete solution is solved
down to truncation errors. Usually it takes 3-5 multigrid iterations to
get the solution to a desired accuracy. Compared to ``exact'' solvers,
as for example FFT, multigrid is an $O(N)$ procedure,
only the multiplicative factor in this estimate could be larger.
Nevertheless,  multigrid is quite comparable in speed to FFT (Foerster
\& Witsch 1982) for medium $N$ and could be more efficient
in large $N$. There exist several versions of parallel multigrid
codes (e.g. R. S. Tuminaro 1989 and G\"artel, Krechel, Niestegge \& Plum 1991).

We compare the exact ($r^{-2}$) force-radius law with that obtained
by the multigrid code in Fig.~1. The figure shows the
averaged pairwise force between a massive particle
and massless test particles that were homogeneously distributed
in logarithmic radial bins and randomly in angular coordinates. The
massive particles were randomly distributed on a finest subgrid.
The force was computed using the usual second order differencing
scheme. The curve labeled TH gives the exact force for point
particles, the curve labeled MG is the usual
smooth-field force that has to go to zero for zero radius in
order to avoid self-forces for a (CIC) particle. The curve labeled
AMG is the result of an adaptive refinement by two levels -- as
we divide the mesh size by 2 for the next finer grid, this curve
is shifted to four times smaller coordinate values, as it
should be.  We have shown also the rms relative radial and tangential force
fluctuations that are around 0.1 per cent of the adaptive mesh force.
The maximum deviations from the mean mesh force that we found in our
experiments were around 10 per cent, but their percentage was
extremely small (only a few cases from an average of about 15000 points
per radial bin). These deviations arose at the boundaries of subgrids
and are typical for the single massive point case;
similar deviations should not occur in the case of a continous
density distribution.

Special care has to be taken with calculation of forces near
boundaries of adaptive refinements. As was mentioned above,
the boundary values of the potential
for local fine grids are obtained by
interpolation from the coarser grid. This guarantees the continuity
of the solution across the boundary, but not necessarily
the continuity of its derivative (force). To smooth
possible jumps in force estimates on boundaries of
subgrids, we found these forces also by interpolation
from the coarser grid.
(We used cubic interpolation
if there were enough data points, otherwise linear interpolation
was used.) As grids are simply a mathematical artefact, particles
should not feel the crossing over between grids of different
resolution. Our procedure enables particles to change
grids smoothly.

Apart from the multigrid Poisson equation solver, other parts of our code
are similar to those in a
usual PM code. We move particles with a standard leapfrog integrator
(we change the timestep, though) on every grid, where the particles are
taken from the point list that
belongs to that grid. The time step is controlled by the Courant condition:
the maximum change of a coordinate should be less than a fraction
of the  mesh size (we have used one-half).
We can have different masses for particles, and for our
cosmological sphere we can guarantee mass conservation by
reflecting the leaving particles back from the opposite side.
The code is not too large, the main modules sum up to about
6200 lines of C.

\section{Testing the code}

The code we developed first is meant to be used for
studying the evolution of generic types of density
singularities. There are only a few of these, all
listed in Arnold (1980), and they could be thought of as
typical progenitors of large-scale structure,
describing the regions of the first collapse.
On scales where there has yet been no significant
interaction of neighbouring elements of structure
(superclusters),
the models of singularities could describe the
actual dynamics and geometry of structure.

In order to understand the evolution of specific
singularities we have to study first the case of
isolated singularities, using vacuum boundary
conditions. In order to minimize the
influence of the geometry of the computational
volume on the results we have to work in a sphere.
Of course, working in a cube would be much simpler,
but we have not been able to get rid of the
ghosts of the cube in the final configurations.

Using isolated boundary conditions on a sphere could
seem to be contrary to the usual cosmological
practice of periodic boundary conditions. In the case
of the tree-code, where vacuum boundaries arise naturally,
people have taken enough trouble to modify the code
to better mimic periodic boundaries (Bouchet \& Hernquist
1988, Hernquist, Bouchet \& Suto 1991).
Thus we have to check if our isolated region models
the evolution of structure in cosmology fairly enough.

As there are really no perfect numerical methods to use
for comparison, one should use the exact solutions for
the evolution of structure. There are only three of
these -- one for the linear regime
and two nonlinear solutions, one for a spherical top-hat
collapse and the other for a one-dimensional plane wave.
The latter solution clearly cannot be used in our case,
so only two remain.

We use the usual cosmological equations for the evolution
of structure for the $\Omega=1$ universe
in comoving cooordinates $\bf x$, connected with
the physical coordinates $\bf r$ by
	\hbox{${\bf r}=a(t)\,\bf x,$}
where
$a(t)$ is the scale factor that describes the expansion
of the universe. We choose this function as the new
time coordinate, and write our basic equations as
$$
	\Delta\Phi=\delta,\eqno(1)
$$
where $Phi$ is a suitably normalized gravitational
potential and $\delta$ is the usual density contrast,
and
$$
	{d{\bf u}\over da}+ {3\over2a}{\bf u} = -{3\over2a^2}{\rm grad}\Phi,
				\eqno(2)
$$
where $ {\bf u}=d{\bf x}/da$.
Our equations are similar to those used by Matarrese, Lucchin,
Moscardini \& Saez (1992).

In order to test for
linear evolution, we have to generate
appropriate initial data --- a realization of a gaussian random
field with a given power spectrum. One starts usually with the power spectrum
of the density contrast:
$$
	P(k)=<|\tilde\delta({\bf k})|^2>_{|{\bf k}|=k},
$$
where $\bf k$ are the wavenumbers and $\tilde\delta$ are the
Fourier amplitudes of the density contrast. We proceed following
the method described by Nusser \& Dekel (1990).

In the linear approximation the movement of particles can be
described by the formula derived by Zeldovich (1970):
$$
	{\bf x}={\bf q}+a\,{\bf u(q)}, \eqno(3)
$$
where $\bf q$ are the initial (Lagrangian) coordinates and
the velocity $\bf u(q)$ depends on $\bf q$ only.
This leads to the (linear) relation between density and
velocity
$$
	\delta({\bf q})=a\,{\rm div}_q\bf u.
$$
The latter relation can be satisfied if
the Fourier amplitudes of the velocity are
$$
	{\bf\tilde u(k)}={{\bf k}\over k^2}\tilde\delta({\bf k}).\eqno(4)
$$

Having chosen the complex Fourier amplitudes for the
density contrast $A({\bf k})+iB({\bf k})$
as random Gaussian numbers with the distribution $N(0,P(k))$
on an appropriate grid $\bf k_i$ in wavenumber space
($-N/2\leq k_i\leq N/2$, where $N$ is the resolution of the grid),
we can form the Fourier amplitudes of the velocity by (4)
and find the velocity field in real space by an inverse
Fourier transform. We use the Fourier transform algoritm for
real 3-D data from Press et~al.\ (1992).

For tests we used a low-resolution model, 28 cells for the
diameter of a sphere (32 for the surrounding cube that we use
to fix the boundary conditions),
in order to clearly see the influence of a discrete grid in
both methods. If we wish
to get a good representation of small perturbations the
density has to be generated from a regular grid.
We chose 8 particles per grid cell which made the total particle
number rather large, $64^3$ for the PM-cube and about 100000
for the sphere.

For the initial density spectrum we chose white noise,
in order to see better the damping of high-frequency
modes. We generated the initial velocity field as described above,
and generated the coordinate displacements from a regular
$\bf q$-grid, choosing them to be
proportional to velocities and normalized to
a fixed (small) rms displacement amplitude. As we work in
a sphere, but the FFT works in rectangular regions, we first find
the displacements in a $32^3$ cube and then set the
displacements outside our sphere to zero.
\beginfigure{2}
	\vskip6.5cm
	\caption{{\bf Figure 2.}
The ratio of the rms density contrast to the value
expected from the linear theory versus the scale factor $a$.
The solid line describes the PM-code, the dotted line stands
for the MG-code.}
\endfigure
\beginfigure{3}
	\vskip6.5cm
	\caption{{\bf Figure 3.}
The normalized rms velocity contrast
(constant in the linear theory) versus the scale factor $a$.
The solid line describes the PM-code, the dotted line stands
for the MG-code.}
\endfigure

We have to take a little more trouble about the velocities.
As the high-frequency modes of the velocity that we
generated cannot be caused (and changed)
by the potential found from the same grid, our
initial state is too hot. To remedy this we
used the quiet start recipe proposed by Efstathiou et~al.\
(1985). We recalculated the velocities, finding the density,
solving for the potential and using the linear approximation
relation between acceleration and velocity, which for our
case ($\Omega=1$) reads:
$$
	{\bf u}={1\over a}{\rm grad}\Phi. \eqno(5)
$$

In the linear case we can check for the evolution of
the velocity and density fields. If the
initial velocities are given by the above formula, it
is easy to see that the velocities have to remain constant,
$ d{\bf u}/dt=0$, (see also (3)). The continuity equation
$$
	{d\delta\over da}+(1+\delta){\rm div}{\bf u}=0
$$
tells us  that the density contrast has to grow linearly with
$a$.
\beginfigure{4}
	\vskip6.5cm
	\caption{{\bf Figure 4.}
The spectral dependence of the ratio of the amplitude
of the density contrast at $a=10$ to its theoretical value.
The absiccae $k$ are given in the units of $1/L$ (the inverse cube
size), the solid curve describes the PM-code and the dotted curve
 the MG-code.}
\endfigure

We started from the rms amplitude 0.025 for
the density contrast and followed the evolution of structure
from $a=1$ until $a=10$. The results are not too good -- see Figs.~2
and 3. In Fig.~2 we show the ratio of our rms density contrast to
the expected value from linear theory (the curve labeled
MG). As is seen, the evolution lags behind the true rate and
the difference reaches about 2.5 times at the scale factor
$a=10$. A similar picture can be seen in Fig.~3 -- while the
rms velocity is expected to remain constant, it actually
drops in time (although the differences are smaller than
these for the density).

This discrepancy is typical for smooth-field simulations,
and is mainly caused by damping of high-frequency modes.
Bouchet et~al.\ (1985) have studied it extensively
in the case of PM-codes. Most of the reasons for the
damping, the CIC density assignment scheme and a finite grid
size, are present in our code too.
In  order to have standard errors to compare with, we solved
the same problems by a standard PM-code. There are, certainly,
better codes around, but simple PM is a pure smooth-field
algorithm, similar to multigrid, in contrast to improved
PPPM-type codes. We used a $32^3$ grid with periodic boundary
conditions, the same number of particles (8) per cell, and
started from the same initial state. The corresponding curves
in Figs.~2 and 3 (labeled PM) look similar, the density contrast
behaves a little better in the PM-code, the velocity drops
a little faster,but differences between the two codes are small.

We can understand what is happening a little better if we look
at the evolution of the density on different scales. In Fig.~4
we give the ratio of the density spectra at the epoch $a=10$
to that at the start of the calculations, normalized by the
theoretical growth factor. If we look at the PM-code curve we
see that the largest scales closely follow the linear growth
law, but the difference comes in at smaller scales. The MG-curve
shows us that this damping is also present in multigrid,
although to a lesser extent. As we calculated our spectra
in a $32^3$ cube, the largest wavelengths present in a cube
do not fit into our sphere and we have a slight drop in
the MG-case these scales, but otherwise the MG-code
with isolated boundary conditions
describes the evolution of small perturbations at least as
well as does the periodic PM-code.

The second test we made is the highly nonlinear spherical
tophat collapse. This is also a textbook problem that can
be solved exactly (see, e.g., Padmanabhan 1993, sec.~8.2).
We shall check for the moment of collapse $a_{coll}$ that is predicted
to be
$$
	a_{coll}=({3\pi\over2})^{2/3}{a_i\over\delta_i},
$$
where $\delta_i$ is the initial density contrast for the sphere
(at the scale $a_i$) and the only approximation used to get
this result is the requirement $\delta_i<<1$.
\beginfigure{5}
	\vskip6.5cm
	\caption{{\bf Figure 5.}
The evolution of the maximum density for a spherical
cosmological collapse. The abcsissae are given in the theoretical
collapse time units,
PM labels the particle-mesh code, MG --- the multigrid
code and ADMG --- the adaptive multigrid code with two additional
finer levels. The ADMG-curve is scaled down (by 15.7) in order to see
all curves together.}
\endfigure

As the dependence of the collapse time on the initial
density contrast is rather strong, one has to take care
when generating the initial distribution. We generated, first,
points on a regular mesh inside a sphere (with an initial grid of
28 cells per diameter), 8 points per cell,
 and left a hollow sphere inside, with
a radius of 0.75 of that of the large sphere.
We then filled this sphere also with points on a regular
mesh, but with a slightly smaller mesh size. In order to
compare the results we generated a similar sphere inside a $32^3$
cube, with a radius of 0.75 of the half-cube size,
 and followed its evolution by our PM-code. Although the
central densities were the same in both cases, the initial
mean densities and density contrasts were different
(the cube had a larger volume) and the predicted collapse times
differ also.

The boundary conditions were fixed (zero for the zero total mass
in a sphere) for the multigrid and periodic for the PM-code.
We estimated the moment of collapse by finding the moment of
maximum density (the density dispersion peaked also at the same
time). Fig.~5 shows the collapse history for both cases,
expressed in the normalized collapse time, $a/a_{coll}$. The simple
multigrid solution (labeled MG) gives a slightly better result
than the PM-code, but they both lag behind the exact
solution. This is due to the smoothing effect of a rather coarse
grid. The PM-peak is higher than the MG-peak, as there is more
mass in the collapsing sphere in the first case.

We solved the problem also with an adaptive multigrid code,
going down two finer levels. Grids on a finer level were
generated when the local density went higher than a chosen
threshold (24 points per cell).
The behaviour of the maximum
density in this case is shown in Fig.~5 by the curve labeled ADMG.
As we see, this gives us a result that is closer to the exact
solution. In this case it is also rather difficult to estimate
the theoretical time -- the CIC scheme gives systematically higher
densities regions
where some components of the density gradient are
zero. This problem can be removed by additional smoothing, but
we did not want to suffer an extra loss resolution.
Instead of this we built the initial CIC density histogram, found
the mass in the central cell at the collapse and estimated the
initial density contrast as an average over this mass.

The adaptive multigrid gets closer to the exact solution, and
the density peak is much better determined here (the ADMG
densities on the graph are divided by 15.7 to get them into
the graph). We looked also at the density distributions, hoping
to see differences, but these were small.

This case allows us also to compare the relative speeds of the
codes. On a SPARCstation-10/51 one timestep of the fully
adaptive multigrid took 5.8 seconds (1.2 seconds for the calculation
of density, 3.6 seconds for the potential, together with new
density calculations at finer grid levels and 1 second to push
the particles). One PM-step took 4.7 seconds ($0.8+0.3+2.3$).
As is seen the time is mainly determined by the number of particles
(106944 for the multigrid sphere, 260488 for the PM-cube) and this
makes them similar; also, the FFT-solver
is about ten times faster for this small grid size.

\section{An application: a high-resolution 3-D pancake}

Careful inspection of published high-resolution cosmological simulations
(in 2-D, of course) indicates that there might be only a few
specific types of elements of structure, and the higher the
resolution is, the more intriguing become the repetitive
structures on smaller and smaller scales. Of course, the
observed supercluster chains and the knots they emerge from
also look similar to some extent.

There is a mathematical basis for this similarity -- if we
agree that visible structure forms in locations of the highest
density (for cold dark matter this means an infinite density),
then we should look for possible classifications of density
singularities. This has been, fortunately, already done,
and the corresponding theory is called the theory
of singularities of Lagrangian mappings. Matter flows in a
gravitating medium follow Lagrangian mappings, so this theory
is relevant to the formation of structure. This was
realized at least ten years ago, and these mappings have been
used in cosmology by Arnold, Shandarin \& Zeldovich (1982).
A very important point is that the number of different
mappings in the generic case (the number of types of structure
elements) is surprisingly small, only six in 3-D space
and four in 2-D (Arnold 1980). If we look at the evolution
of these singularities in time, we are dealing with a metamorphosis
of the singularities, and there are from two to five types of
evolution for every basic singularity type, which is a
small number.

The singularity mappings describe the motion of matter
 only until flows intersect and
it is not clear how useful they are afterwards. And as these
mappings are local, we do not know how long they will remain so
before being distorted by interaction with neighbouring
singularities.

This all is a subject of fascinating study, and we can use the
basic types of mappings to find the initial conditions for
the emerging structure. As they are generic, these are the
structures that must be most common both in the sky and in
the simulations. The code that we can use to study the
formation of structure has clearly to be as high-resolution
as possible. This was the motive for starting the development
of the present code.

As usual in a new field, there is a mass of new problems and
intricacies here. We shall demonstrate how our code works
and shall describe these problems using the most familiar
Lagrangian singularity -- the Zeldovich pancake. It belongs
to so-called type $A_3$, and the birth of a pancake is
described by the metamorphosis $A_3(-,+)$.

This mapping can be described by the coordinate transformation
(from Lagrangian coordinates to the normal Eulerian space)
$$
\eqalign{
	x_1&=4q_1^3+2(q_2^2+q_3^2-t)q_1,\cr
        x_2&=q_2,\cr
        x_3&=q_3,\cr}\eqno(6)
$$
where $\bf x$ are the Eulerian and $\bf q$  the Lagrangian
coordinates. This mapping is meant to be used near the
zero point of coordinates, that is the place where the
pancake is born -- the density is
$$
	\rho({\bf x(q)})=|{\partial{\bf x}\over\partial{\bf q}}|^{-1}=
		|12q_1^2+2(q_2^2+q_3^2-t)|^{-1},
$$
and, when time grows from $-\infty$ towards the future,
this density first becomes infinite at $t=0$ in $\bf x=0$.

The velocities for this mapping can also be found
(Suisalu 1987):
$$
\eqalign{
	v_1&=-2q_1,\cr
	v_2&=0,\cr
	v_3&=0\cr}\eqno(7)
$$
(all $A$-type singularities are essentially one-dimensional).

These formulas are in principle all that one needs to set up
a pancake birth simulation. The remainder are technical details,
but they are rather important. The main problem is that the
mapping is nonlinear, and if we want to model it, we must
restrict the mapping to a finite region -- a sphere is the best
choice, as it minimally distorts the final results. The mappings
(6) and (7) are free to change using any diffeomorphism we want,
as this does not essentially change the mapping in the centre of the
coordinates. However, if we want to study as large a region around
the centre as possible, the modification must be minimal.
Our choice is
$$
	{\bf x=f(q)}(1-q^2/R^2)+{\bf q}\,q^2/R^2,\eqno(8)
$$
where $\bf f(q)$ is the mapping (6) and $R$ is the radius
of the sphere. The velocities are changed similarly:
$$
	{\bf v=w(q)}(1-q^2/R^2),\eqno(9)
$$
where $\bf w(q)$ is the original (Eulerian) velocity in the
Lagrangian coordinates (7). This is not a perfect solution,
as it gives us a constant density and zero velocities
on a Lagrangian sphere $q^2=R^2$, and thus distorts the
geometry of our Eulerian computational volume. We have fought this by
choosing a small radius, $R=0.05$, as the region inside it
maps practically into
an Eulerian sphere (our initial time parameter $t=0.4$).
For simulations we choose the scale factor
$a$ as the time variable. The times $a$ and $t$ can be connected to
each other by any monotonic transformation.
We are also at liberty to choose the velocity amplitude -- this
amounts to changing the time unit. We did this using the quiet
start recipe -- we built the point distribution by the mapping (8),
found the density (the maximum initial density contrast was 0.308),
 solved for the potential and used the linear
evolution formula (5) to find the initial dynamical velocities.
We used both the dynamical velocities and the velocities from
the mapping scaled down to make the maximum velocities coincide
in both cases.
\beginfigure*{6}
	\vskip19.5cm
	\caption{{\bf Figure 6.}
The central $x_1$-$x_2$ slices of the simulation of the
birth of a pancake ($A_3(-,+)$ metamorphosis) at the epoch $a=13.6$.
The simulation started at $a=1$ with a maximum density contrast
$\delta_{max}=0.308$ and with the velocity amplitude found from
linear dynamics.
Panels (a--d) show the particle distribution in
a slice on different scales; the boxes show adaptive subgrids.}
\endfigure

The initial grid is the same as we
used for the spherical collapse,
a sphere with diameter of 28 cells (in Lagrangian
coordinates). In order to better see the details of the structure
we choose 27 points per grid cell, distributed regularly,
a total of 324609 points.
If the number of points per cell gets larger than 70
we create a finer rectangular subgrid with double linear resolution.
We have limited the number of refinement levels to five --
this limit is imposed mostly by the noise that accumulates
during the run. This means that the effective spatial resolution
in the central subgrid is 1/1024. Also, in this case the
boundary conditions will
change in time -- we found them by direct summation over the
initial ($32^3$) grid. We use the same equations of motion
as before, working in the $\Omega=1$ cosmology.
\beginfigure*{7}
	\vskip19.5cm
	\caption{{\bf Figure 7.}
The central $x_2$-$x_3$ slices of the same density distribution
as in Fig.~6.
Panels (a--d) show the particle distribution in
a slice on different scales; the boxes show adaptive subgrids.}
\endfigure

We illustrate the results by a series of figures.
As the results for the initial velocities from the
mappings and from the dynamics did not differ much,
we shall use the case of the mapping (truly unidirectional)
velocities. The figures all refer to the time $a=13.6$
(we started with $a=1$). The first series of figures (Fig.~6a-d) show
the distribution of mass points in a thin (one cell of the
basic grid) slice along the $x_1$-$x_2$ coordinate plane. We have
also shown the borders of the subgrids, and
the change of scale can be followed from the coordinate values.
\beginfigure*{8}
	\vskip19.5cm
	\caption{{\bf Figure 8.}
A three-dimensional representation of the same density
distribution as in Figs. 6 and 7. A region of space between
chosen density levels is cleared to show the inner density
levels (see Table~1). Panels (a--d)
show the density distribution
in increasingly smaller scales - the small cube in the centre
shows the size of the next panel. }
\endfigure

As this is already a well advanced state of collapse, we
see a fairly rich substructure in the figures. Fig.~6a
covers the whole slice, and the pancake is clearly seen in the centre.
The edges of the pancake are formed by the turnback points
of the first particle stream that has passed the pancake
plane. Only four subgrids can be seen in this figure, the two
smallest are too small. The lenselike overall
geometry of the collapse is not caused by its living in a sphere,
this is a generic property of this type of singularity and is
determined by the mapping itself. It is possible to distort
the shape a little, in principle, but we discovered that such
a distortion will not live long --- the mapping we used is generic.

Fig.~6b (coinciding with the third subgrid)
shows the central part of the collapsed region --
we see two perpendicular planes and the formation of ellipsoidal
shells -- these are the turnback regions of smaller-scale
flows, and the collapse tends to become more spherical. This
differs from the picture seen in high-resolution 2-D simulations
by Beacom et~al.\ (1991), and
is probably caused by the fact that 3-D gravitation is in general
more effective than 2-D gravitation. These shells cannot be caused
by the spherical boundary, these flows live in the centre where
they do not feel the the large-scale symmetry. They cannot either
be the result of the growing temperature, this effect can be seen
in the centre only (Fig.~6d).

Fig.~6c shows the central matter distribution in more detail --
we see a much smaller pancake in the centre, with a size
about 80 times smaller than that of the largest pancake.
The last figure in this series, Fig.~6d shows no more detail
although the last grid has $13\times17$ cells in this plane.
This is probably caused by the (numerical) heating during the collapse.
The central density contrast is $1.6\cdot 10^5$, this value
can be taken to characterize the dynamical resolution of the
simulation.

Figs.~7a-d show a slice from the $x_2$-$x_3$ plane, using similar
scales as the previous series. In Fig.~7a we see the whole plane,
where the outer density enhancement is the edge of the main
pancake. More inner density ridges can be seen in Fig.~7b. They
are rather well resolved in Fig.~7c, and the central part can be
seen as a small hot lens at Fig.~7d. The spokes along the
coordinate axes that are evident in all these figures are either
caused by the anisotropy of the \hbox{7-point} difference operator,
or by the CIC density assignment algorithm that gives enhanced
densities along the coordinate planes and axes.
These spokes are not too prominent, however.

In the last series of figures (Fig.~8a-d) we have tried to show
the 3-D density distribution. Each of these figures is
a 3-D representation of three density levels, cut in half
by an \hbox{$x_1$-$x_2$} coordinate plane (in these figures the
coordinate $x_1$ is vertical). The smaller inner cube
in Figs.~8a-c shows the size of the large cube in the
next figure. The density values for the different level
surfaces (the outer one, the inner surface of the ``shell''
and the outer surface of the central detail) are given
in Table~1. The surfaces are better seen in Fig.~8a, where
the density resolution is $1/32$ of the cube size. In Figs.~8b-d
the resolution is $1/24$ and is probably a little too low
for the IDL(Interactive Data Language -- a graphical software
package from Research Inc.) shade-volume command to manage. Of course, in
the real density distribution there are no holes, they are
cut out in the figures only to help to visualize the
continuous density distribution.

\begintable{1}
	\caption{{\bf Table 1.} Density levels in the 3-D configuration}
\halign{\tabskip20pt\hfil#\hfil&\hfil#&\hfil#&\hfil#\cr
\noalign{\smallskip\hrule\smallskip}
panel&\multispan3\hfil\hbox{density level}\hfil\cr
&outer&inner&central\cr
\noalign{\smallskip\hrule\smallskip}
a&0.02&0.7&10.5\cr
b&14.4&33.5&173.3\cr
c&185.2&924.4&11090.0\cr
d&9244.0&46220.0&154000.0\cr}
\endtable

Fig.~8a starts showing us the basic sphere and the primary
pancake inside. The outer shell is, in fact, continuous, and
looks striped by the IDL efforts. Fig.~8b shows a second pancake
inside the first one (look at the sizing cubes) and a high-density
detail at the centre. This detail is better resolved in Fig.~8c,
showing a density enhancement that is oriented perpendicularly
to the original pancake plane (there is a trace of it in Fig.~6b,
the horizontal density enhancement). The plane itself has also
rather high density here.

The central core of Fig.~8c is resolved in Fig.~8d -- it is
a lenselike density concentration along the original
pancake plane (the vertical density enhancement in Fig.~6b),
and the small pancake of Fig.~6d lives in its centre.

As is seen, the inner regions grow more and more irregular,
the smaller the scales and the larger the densities. This
could be due to a number of reasons. The first of them
is the numerical heating caused mainly by the force
fluctuations at the edges of subgrids. Another is the fact
that we have probably not taken proper care when arranging
the initial mappings, and as a result these are not cool
enough.

\section{Conclusions}

We have presented here an adaptive multigrid code for
gravitational simulations and tested it for
cosmological problems. The multigrid approach lends
itself naturally to adaptive refinement and does not impose any
restrictions on the boundary conditions to be used. It
does not much use more memory than the popular cosmological
codes and it is fast enough to be used on present computers.

The list of possible enhancements is rather long.
We have already implemented the case of periodic boundary conditions
that are more suitable to simulate the evolution of global
structure. We are planning to use better difference operators
in order to get more isotropic forces, and we have ideas on how
to improve the density assignment algorithm. The speed of
our algorithm is still lower than that of the FFT, but we
can use the potential and grids from the previous time step
to enhance the convergence -- this can be done because
we keep the potential separate
from the density field. Probably the use of block time will
also speed up the code. We determine our boundary conditions
at present by direct summation over the grid -- we can either use
a coarser grid for this or use a FFT solver.
I.S. has implemented the code also on
a parallel computer (2-D case on a CM-200), but this implementation
needs further work.

As for the present
application of following the structure of singularities, this
has shown us the importance of setting up clean initial
conditions and getting rid of noise for truly high-resolution
simulations. A typical example is the CIC density assigment
scheme -- the fact that it may give spurious density enhancements
does not worry anybody so far as we are using noisy initial
conditions, but in the present case its deficiencies were
obvious.

\section*{Acknowledgements}

This code has taken a long time to mature, and I.S. has
worked on it, besides at his home institute, Tartu Astrophysical
Observatory, at ESO, Cambridge, at MPI f\"ur
Astrophysik, but longest at NORDITA, where this work was
finished. E.S. spent a month at NORDITA to help to bring
the code into the cosmological environment.  We thank all these
institutions for their hospitality and stimulating atmosphere.

We thank also our referee, Dr. H.M.P. Couchman, for his competent
and useful advice.

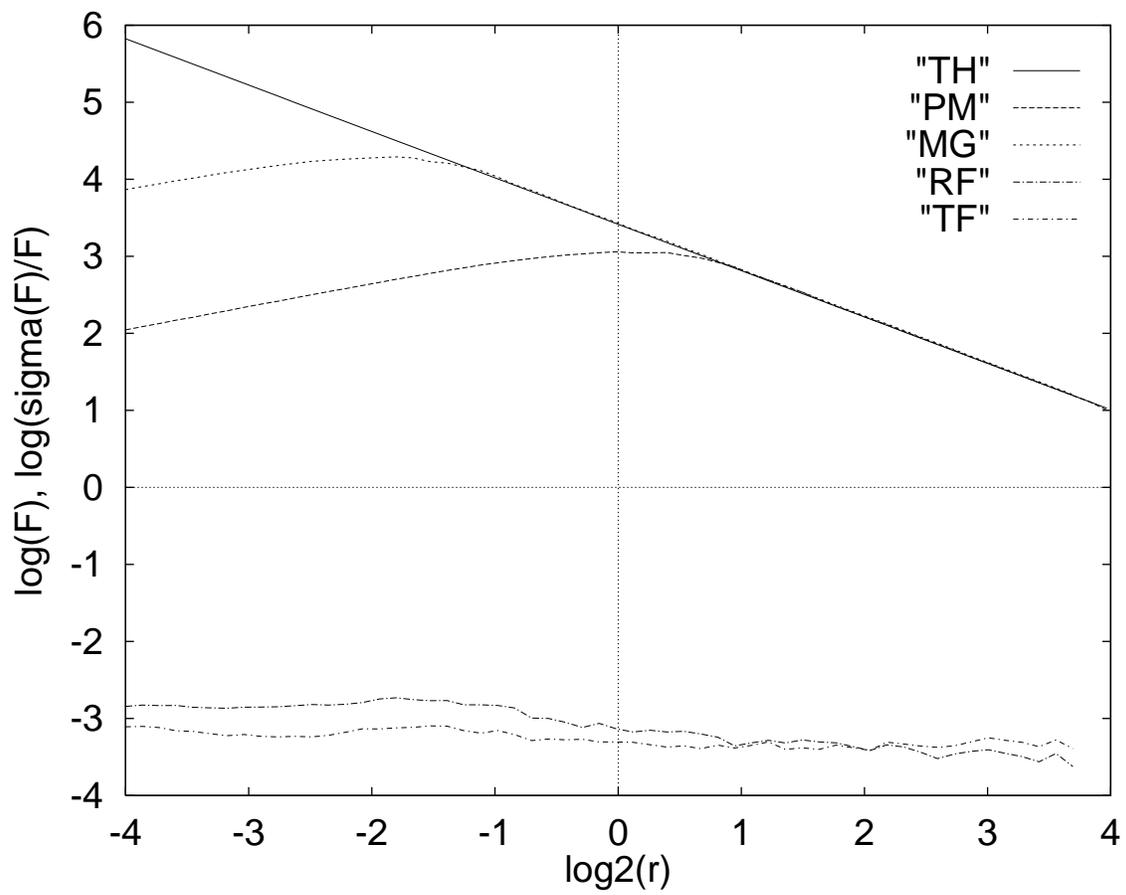
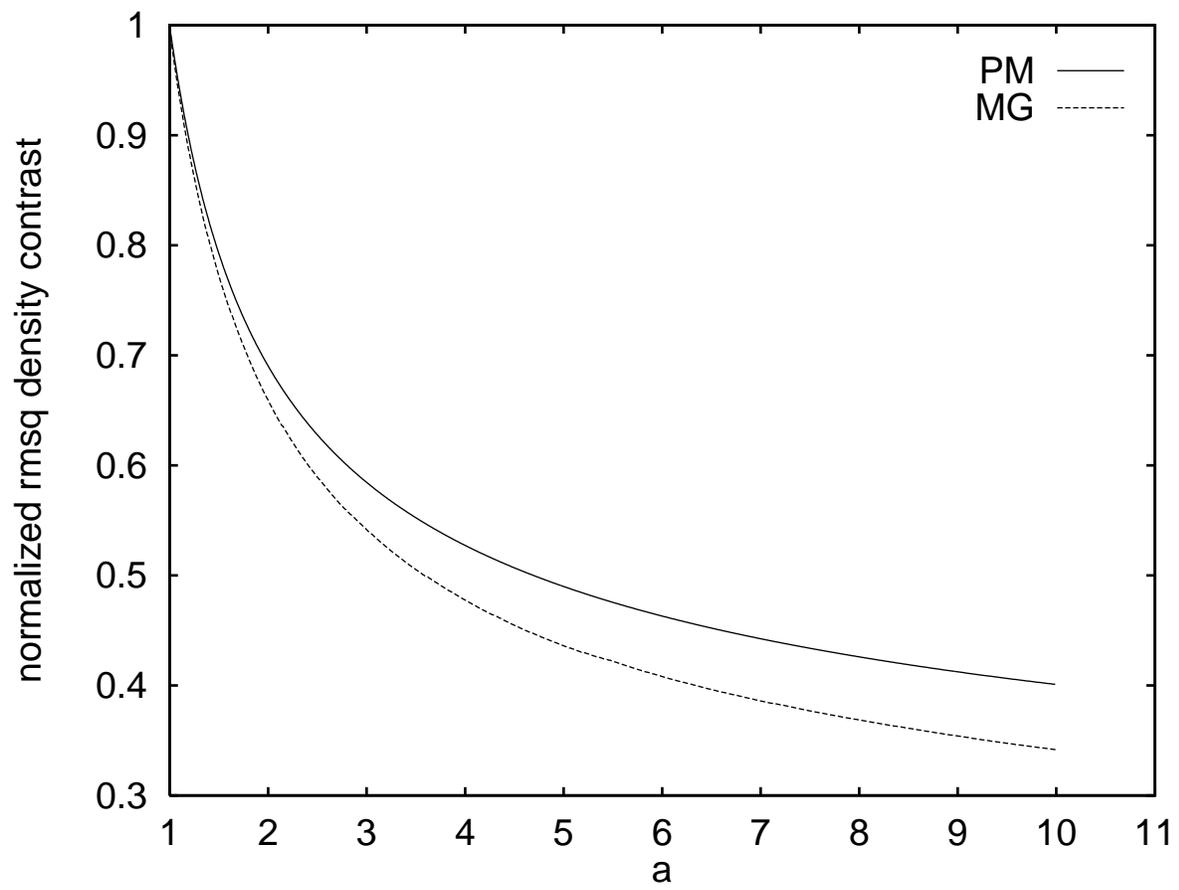
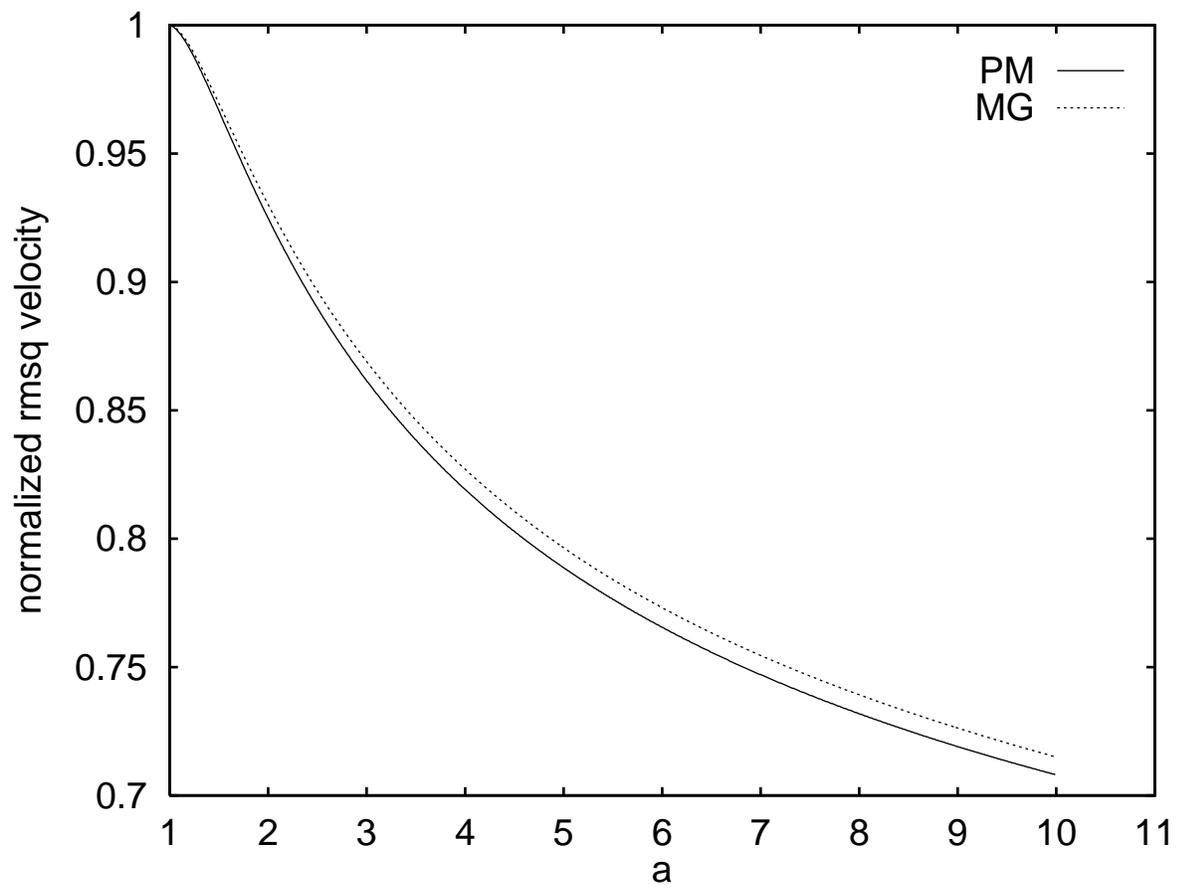
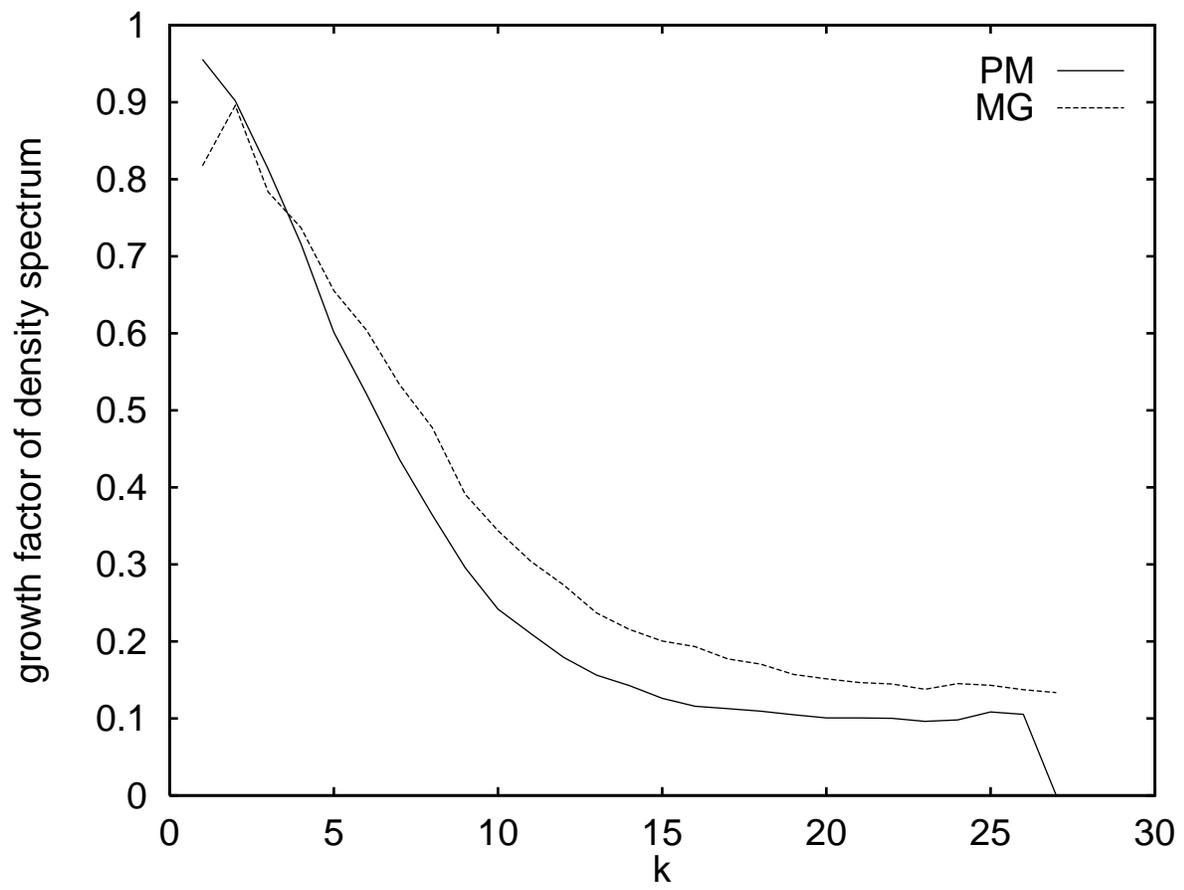
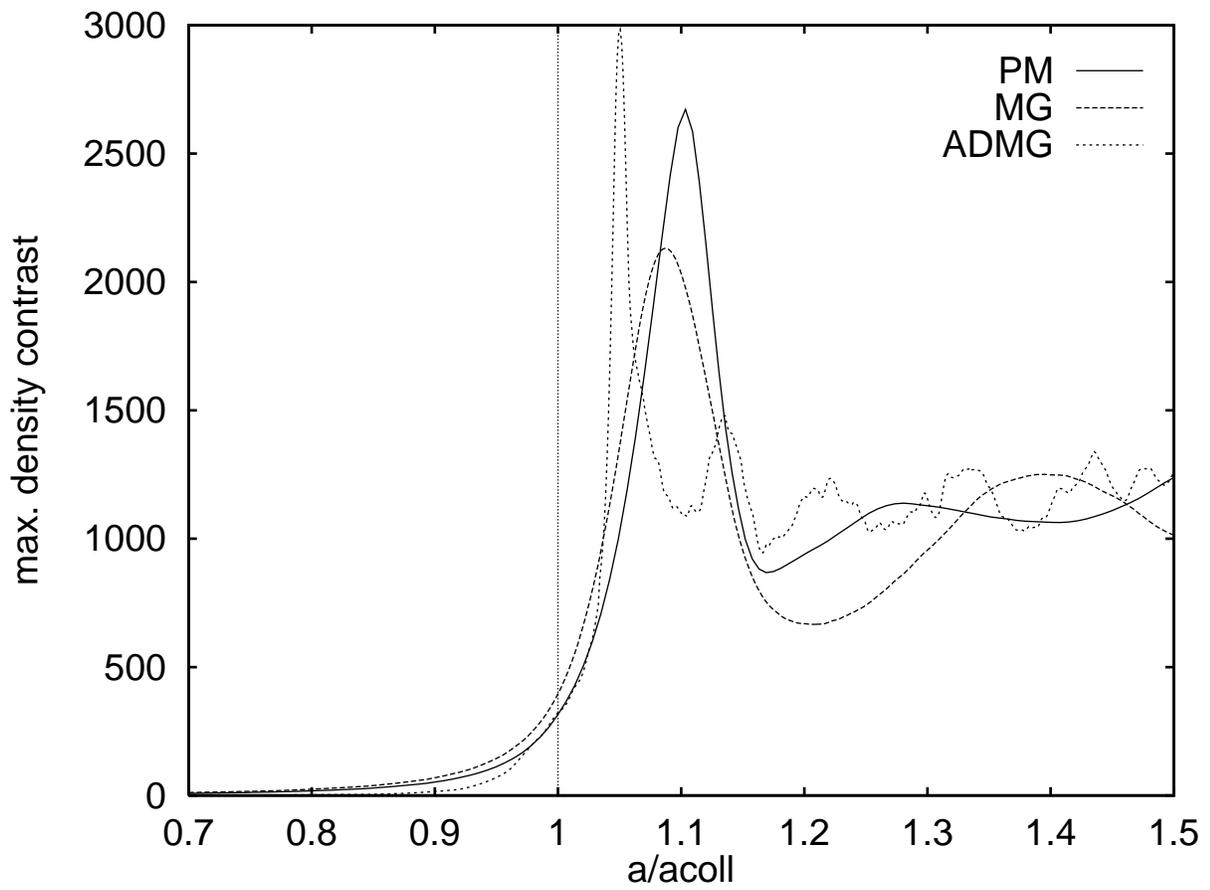
\bye